\documentclass[conference]{IEEEtran}
\IEEEoverridecommandlockouts

\usepackage{cite}
\usepackage{amsmath,amssymb,amsfonts}
\usepackage{braket}
\usepackage{booktabs}
\usepackage{amsthm}
\usepackage{algorithmic}
\usepackage{graphicx}
\usepackage{textcomp}
\usepackage{comment}
\usepackage{xcolor}
\usepackage{color, soul}
\usepackage{tikz}
\usepackage{verbatim}
\newcommand\copyrighttext{%
  \footnotesize \textcopyright 2026 IEEE. Personal use of this material is permitted.
  Permission from IEEE must be obtained for all other uses, in any current or future
  media, including reprinting/republishing this material for advertising or promotional
  purposes, creating new collective works, for resale or redistribution to servers or
  lists, or reuse of any copyrighted component of this work in other works.}
\newcommand\copyrightnotice{%
\begin{tikzpicture}[remember picture,overlay]
\node[anchor=south,yshift=10pt] at (current page.south) 
  {\fbox{\parbox{\dimexpr\textwidth-\fboxsep-\fboxrule\relax}{\copyrighttext}}};
\end{tikzpicture}%
}
\def\BibTeX{{\rm B\kern-.05em{\sc i\kern-.025em b}\kern-.08em
    T\kern-.1667em\lower.7ex\hbox{E}\kern-.125emX}}
    
\begin{document}

\title{
Formally Verifying Quantum Phase Estimation Circuits with 1,000+ Qubits
}
\author{\IEEEauthorblockN{Arun Govindankutty and Sudarshan K. Srinivasan}
\IEEEauthorblockA{\textit{Electrical and Computer Engineering Department, North Dakota State University, Fargo, ND, USA} \\
}}
\maketitle
\copyrightnotice
\begin{abstract}
We present a scalable formal verification methodology for Quantum Phase Estimation (QPE) circuits.  
Our approach uses a symbolic qubit abstraction based on quantifier-free bit-vector logic, capturing key quantum phenomena, including superposition, rotation, and measurement. The proposed methodology maps quantum circuit functional behaviour from Hilbert space to a bit-vector domain.  
We develop formal properties aligned with this abstraction to ensure functional correctness of QPE circuits.  
The method scales efficiently, verifying QPE circuits with up to 6 precision qubits and 1,024 phase qubits using under 7.5~GB of memory.
\end{abstract}

\begin{IEEEkeywords}
formal verification, quantum computing, quantum circuit verification, quantum phase estimation. 
\end{IEEEkeywords}

\section{Introduction}
\label{sec:intro}

Quantum computing has demonstrated the potential to surpass classical computation in several classes of problems across domains such as communication, machine learning, materials science, medicine, and optimization, enabling progress toward quantum advantage~\cite{IET_futureQC, qc_arun, tqe_portfolio}. 
Central to many of these advances are quantum algorithms that address problems intractable for classical systems. 
Among them, Quantum Phase Estimation (QPE) is a core primitive, employed in key applications including Shor’s algorithm~\cite{shor_algo}, quantum counting and amplitude amplification~\cite{quantum_counting}, eigenvalue estimation~\cite{quant_eigen_value}, and quantum linear system solvers~\cite{linear_eqn}. 
Realizing these applications in practice requires QPE circuits to scale to thousands of qubits. 
However, increasing circuit scale also amplifies design and functional complexity, significantly raising the risk of implementation errors. 
Therefore, establishing the functional correctness of large-scale QPE circuits is a critical requirement for reliable quantum computation.

Quantum circuits operate over complex Hilbert spaces and exhibit uniquely quantum effects such as superposition, entanglement, and measurement, making functional verification inherently challenging. 
Formal verification is well suited to uncover subtle design errors while ensuring design correctness~\cite{formal_default}. 
Recent work has extended these techniques to the quantum domain by adapting classical formal methods to reason about quantum program behavior~\cite{quantum_formal_survey, arun_review}.

This work explicitly incorporates measurement into the abstraction framework to the work presented in~\cite{iqft_arun}, yielding a generic symbolic qubit abstraction based on quantifier free bit-vector logic. 
The resulting model enables functional correctness verification using Satisfiability Modulo Theories (SMT) solvers. 
Our main contributions are as follows:
\begin{itemize}
    \item A unified qubit abstraction integrating rotational, superposition, and measurement operations.
    \item Symbolic abstract modelling of controlled modular exponentiation in QPE circuits.
    \item Formal specification of correctness properties capturing QPE functionality.
    \item Soundness lemmas and formal proofs of the proposed verification methodology.
\end{itemize}
The required theoretical background is presented next.

\section{Background}
\label{sec:background}

This section briefly reviews qubits, quantum gates, and Quantum Phase Estimation (QPE). Detailed treatments can be found in~\cite{qc_textbook, qc_book_mit, qc_book_LaPierre2021}. 
The basic unit of quantum information is the qubit, represented as a linear combination of the computational basis states $\ket{0}$ and $\ket{1}$, defined as
\begin{equation}
\ket{0} =
\begin{bmatrix}
1\\
0
\end{bmatrix}, \quad
\ket{1} =
\begin{bmatrix}
0\\
1
\end{bmatrix}.
\end{equation}

\begin{figure}[!h]
\centering
\includegraphics[width=0.5\textwidth]{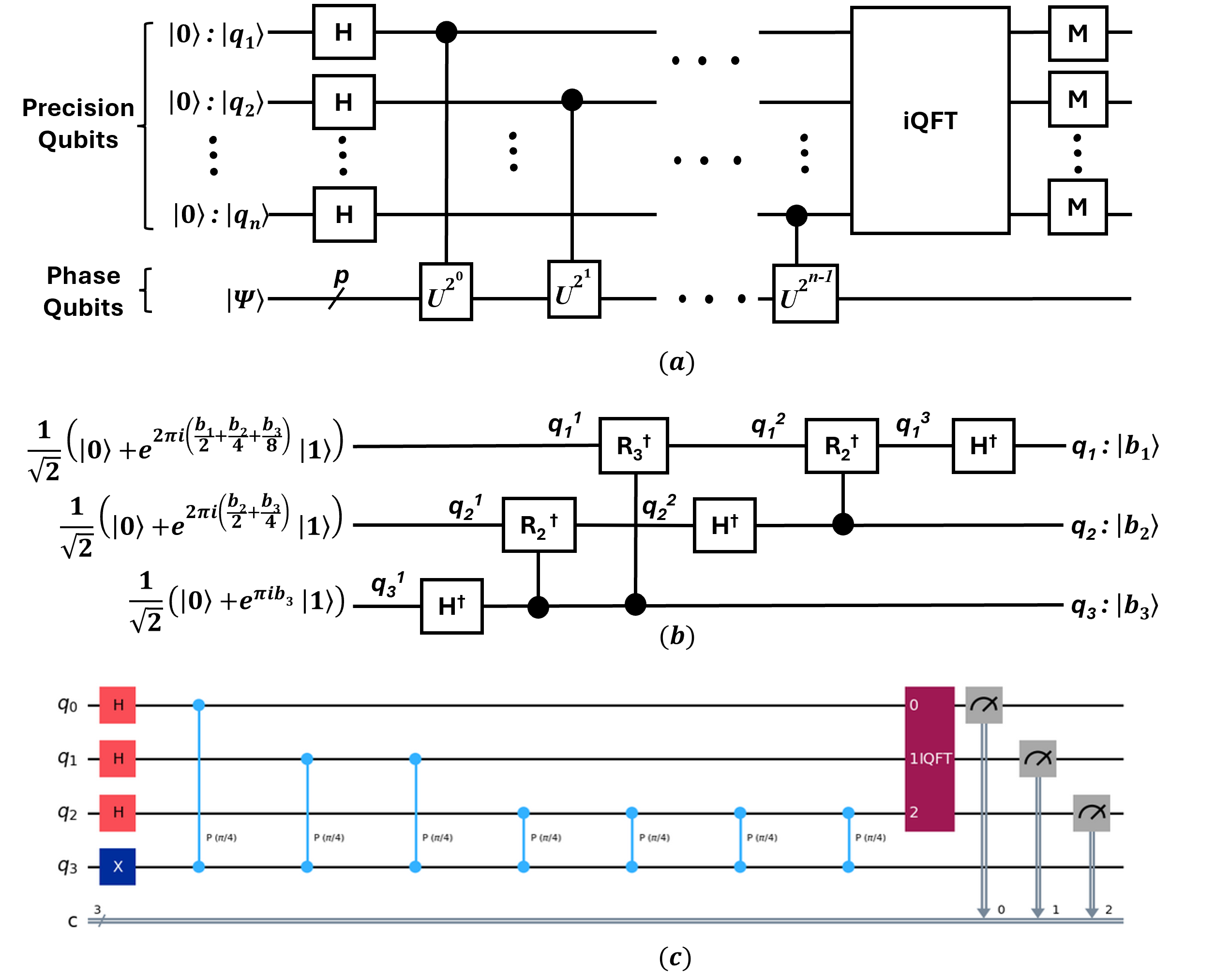}
\caption{(a) Quantum phase estimation circuit. (b)3-qubit inverse quantum Fourier transform circuit~\cite{iqft_arun}. (c) Quantum phase estimation circuit with 3 precision qubits and 1 phase qubit implemented in IBM-Qiskit.}
\label{fig:ckt_fig_qpe}
\end{figure}

Quantum gates are unitary operators that act on qubits to transform their states. 
Given a unitary operator $U$, an eigenstate $\ket{\psi}$, and $n$ precision (ancilla) qubits, a QPE circuit estimates the phase $\theta$ with precision $1/2^n$, where
\begin{equation}
U\ket{\psi} = e^{2\pi i \theta}\ket{\psi}.
\end{equation}

Figure~\ref{fig:ckt_fig_qpe}(a) shows a standard QPE circuit~\cite{qc_textbook}. 
The circuit consists of Hadamard (H) gates to create equal superposition, controlled modular exponentiation operators C-$U^{2^k}$, the inverse Quantum Fourier Transform (iQFT), and measurement (M). 
As illustrated in Figure~\ref{fig:ckt_fig_qpe}(b), the iQFT is composed of inverse Hadamard ($\text{H}^\dagger$) and inverse controlled-rotation (C-R$^\dagger_k$) gates. 
The specific form of $U$ (in C-$U^{2^k}$) depends on the target application. In this work, we assume a generic unitary operator with $p$ qubits.
\begin{equation}
H = H^\dagger = \frac{1}{\sqrt{2}}
\begin{bmatrix}
1 & 1\\
1 & e^{\pi i}
\end{bmatrix} = \frac{1}{\sqrt{2}}
\begin{bmatrix}
1 & 1\\
1 & -1
\end{bmatrix}
\end{equation}
\begin{equation}
\text{iC-R}_k^\dagger = 
\begin{bmatrix}
1 & 0\\
0 & e^{-2 \pi i/2^k}
\end{bmatrix}
\end{equation}

A Quantum Phase Estimation (QPE) circuit consists of $n+p$ qubits. 
The first $n$ \emph{precision qubits} control the estimation accuracy, while the remaining $p$ \emph{phase qubits} encode the system state $\ket{\psi}$ whose phase is estimated. 
Controlled modular exponentiation operations (C-$U^{2^k}$) are applied to the phase qubits, followed by the inverse Quantum Fourier Transform (iQFT) on the precision qubits and final measurement.
Ensuring functional correctness of QPE requires verifying each stage of the circuit, including superposition of the precision qubits, controlled modular exponentiation, iQFT, and measurement. 
Accordingly, our methodology proceeds in two steps: (i) verification of the precision qubits and (ii) verification of the phase qubits.

\section{Existing Approaches}
\label{sec:rel_works}
This section reviews work most closely related to our methodology.
Hietala \textit{et al.}~\cite{sqir_2021} proposed SQIR, a Coq-based intermediate representation enabling machine-checked proofs of quantum algorithms such as QPE. 
SQIR verifies algorithmic specifications and supports extraction to executable languages (e.g., QASM~\cite{qasm_cite}, Qiskit~\cite{qiskit}). 
However, it does not verify implemented circuits, requires substantial formalization (over 3,700 lines), and does not report scalability. 
In contrast, our approach verifies QPE implementations using only four concise properties and scales to 1,024 qubits with 6-bit precision.
Chareton \textit{et al.}~\cite{deductive_verif} introduced Qbricks, a deductive verification framework based on symbolic path-sum semantics. 
While highly automated, scalability results are not reported. 
Our method instead applies abstraction to reduce quantum verification to bit-vector reasoning, enabling efficient and scalable verification of large QPE circuits.
Peng \textit{et al.}~\cite{verif_shor} presented a fully mechanized Coq verification of Shor’s algorithm using reversible circuits (RCIR). 
Although demonstrating end-to-end algorithmic correctness, the approach incurs significant formal effort and does not address scalability or verification of implemented circuits. 
Our methodology directly targets implemented QPE circuits~\cite{qft_arun,qenc_arun} and supports optimized designs while maintaining scalability.
Several works apply formal verification to quantum circuits using equivalence checking~\cite{yamashita2010, Feng2013}, path-sum representations~\cite{qc_large_verif}, and theorem proving~\cite{liu2019}. 
Unlike these approaches, our method does not require a golden reference, avoids arithmetic overflow inherent to dyadic path-sum techniques, and offers superior scalability. 
We address these limitations and present a scalable formal verification of complete QPE circuits.

\section{Abstraction Framework}
\label{sec:abs}
This section introduces the proposed abstraction framework by formally defining abstractions for qubits, gates, and operators. 
The framework exploits the fact that a global phase does not affect observable quantum behaviour~\cite{qc_textbook}. 
Any qubit state can thus be expressed using a relative phase rotation on one basis state, as
\begin{equation}
\ket{\psi} = e^{\phi_1}\alpha\ket{0} + e^{\phi_2}\beta\ket{1}
           = e^{\phi_1}\big(\alpha\ket{0} + e^{\phi_2-\phi_1}\beta\ket{1}\big).
\end{equation}

Accordingly, our qubit abstraction retains only a single rotational component, capturing the relative phase term $e^{\phi_2-\phi_1}$ as a rotation applied to $\ket{1}$. 
This representation can be decomposed into independent rotations on $\ket{0}$ and $\ket{1}$, denoted $r_{\ket{0}}$ and $r_{\ket{1}}$, following~\cite{iqft_arun}. 
The framework further supports application-driven abstraction simplification. 
We next formalize the proposed qubit abstraction.\\

\noindent
\textbf{Definition 1.} \textit{(Abstract Qubit)}
\textit{The bit-vector abstraction Q of a qubit is a 4-tuple 
$\langle q, s, r, m\rangle$,
where $q, s$, $r$, and $m$
are bit-vectors of length \textup{1}, $\lceil log_2(\max(H)\text{+2})\rceil$, $n+p$, and \textup{2}, respectively.\\}

\noindent
The abstract qubit consists of four components: the basis state $q$, the superposition component $s$, the rotational component $r$, and the measurement state $m$.
The basis state $q$ is encoded using a single bit since a qubit outside superposition lies in $\{0,1\}$.
The component $s$ captures the superposition state of the qubit and tracks the number of Hadamard ($H$) gates applied to the qubit. Its length depends on the circuit and is determined by $\max(H)$, the maximum number of $H$ gates on any qubit.
This information is critical for verification, as an odd value of $s$ indicates that the qubit is in superposition, and even, not.
For example, in QPE, $\max(H)=3$, yielding $\text{length}(s)=3$.

The component $r$ captures symbolic phase rotation and has length equal to the total number of qubits in the circuit.
In QPE, phase accumulation from the inverse Quantum Fourier Transform (iQFT) can be represented as symbolic rotations applied to the $\ket{1}$ state. Phase induced by controlled modular exponentiation (C-$U^{2^k}$), reduced as relativistic phase rotations and are captured by symbolic rotations of $r$ by C-$U$ gate.

The abstraction of each quantum gate is defined next. 
The Hadamard ($H$) gate transforms an abstract qubit
$\langle q^i, s^i, r^i, m^i \rangle$
to
$\langle q^o, s^o, r^o, m^o \rangle$,
where superscripts $i$ and $o$ denote input and output states.\\

\noindent
\textbf{Definition 2 (Abstract $H$ Gate).}~\cite{iqft_arun}
\textit{If $s^i=\max(H)+2$, then $s^o \leftarrow \max(H)+1$; otherwise $s^o \leftarrow s^i+1$.
If ($s^i$ is even) $\wedge$ $q^i=\text{1b0}$, then $q^o=q^i$ and $r^o \leftarrow r^i$;
else if ($s^i$ is even) $\wedge$ $q^i=\text{1b1}$, then $q^o=q^i$ and $r^o \leftarrow r^i +_n \text{nb}10\ldots0$;
else if ($s^i$ is odd) $\wedge$ $q^i=\text{1b1}$, then $q^o=r^i[n]$ and $r^o \leftarrow r^i -_n \text{nb}10\ldots0$;
else if ($s^i$ is odd) $\wedge$ $q^i=\text{1b0}$, then $q^o=r^i[n]$ and $r^o \leftarrow r^i$.\\}

\noindent
Here, $+_n$ and $-_n$ denote fixed-point modular addition and subtraction with respect to $n$. 
Since QPE employs the iQFT, the abstraction must support both QFT and iQFT. 
Accordingly, $\max(H)=1$ for QFT and $\max(H)=2$ for iQFT and QPE, yielding a superposition bit-vector of length three for QPE (Definition~1).

The $H$ gate toggles $s$ by incrementing it. For length($s$)=3, 3b000,$\dots$,3b111. After 3b111, value of $s$ toggles between 3b110 and 3b111. Values exceeding $3\text{b}010$ in QPE indicate an error. 
Phase is captured by updating the rotational component $r$ using modular arithmetic: entering superposition applies a clockwise rotation via $+_n \text{nb}10\ldots0$, while exiting superposition applies an anti-clockwise rotation via $-_n \text{nb}10\ldots0$. 
During superposition entry, $q^o=q^i$, and during exit, $q^o$ is assigned the most significant bit of $r^i$. 
Since $H$ is self-inverse ($H^\dagger=H$), the abstraction is valid in both directions. 
The $H$ gate affects only $q$, $s$, and $r$, leaving the measurement component unchanged ($m^o=m^i$).
The abstraction of the inverse controlled-rotation (C-R$^\dagger_k$) gate is presented next.\\

\noindent
\textbf{Definition 3.} \textit{(Abstract inverse Controlled-Rotation (C-R$^\dagger_k$) Gate)}~\cite{iqft_arun}
\textit{If ($q^i_{c}$=1b1 $\wedge$ $s^i_{c}$=3b010 $\wedge$ $s^i_{t}$=3b001), 
then $r^o_t \leftarrow r^i_t -_{n} \langle kb00..01_{k-n}0...0\rangle$,
else  $r^o_t \leftarrow  r^i_t$,  
$q^o_{t} \leftarrow q^i_{t}$, 
$s^o_{t} \leftarrow s^i_{t}$.\\ 
 } 
 
\noindent
Inverse controlled-rotation (C-R$^\dagger_k$) gates appear in the iQFT stage of QPE circuits. 
As per Definition~1, their abstraction uses a superposition bit-vector of length $3$, derived from $\lceil \log_2(\max(H)+2) \rceil$. 
The abstract C-R$^\dagger_k$ gate operates on a control qubit $Q_c$ and a target qubit $Q_t$. 
When $Q_c$ is in state $\ket{1}$ and not in superposition, and $Q_t$ is in superposition, the gate induces an anti-clockwise phase rotation of $2\pi/2^k$ on $Q_t$. 
This effect is captured by modular subtraction from the rotational component $r_t$, encoding the rotation magnitude symbolically. 
In all other cases, the abstraction leaves the qubit state unchanged. 

The C-R$^\dagger_k$ gate updates only the rotational component, while the basis, superposition, and measurement components remain unchanged. 
The corresponding clockwise controlled-rotation (C-R$_k$) gate is modelled analogously using modular addition, but is omitted here since QPE circuits employ only C-R$^\dagger_k$ gates. 
The abstraction of the controlled unitary (C-$U$) operation is presented next.\\

\noindent
\textbf{Definition 4.} \textit{(Abstract controlled-unitary (C-$U$) Gate)}
\textit{ $\forall \ l \in\{1, 2, \dots, p\}$, if $q_c^i$=1b0, then, $r^o_{t_l} \leftarrow r^i_{t_l}$, else if 
$q_c^i$=1b1, then, $r^o_{t_l} \leftarrow r^i_{t_l} +_2 \ r_l $.\\
 }

\noindent
A controlled unitary operation conditionally applies the unitary operator $U$ within the QPE circuit depending on the basis state of the precision qubits. The operation takes a precision qubit acting as the control ($Q_c$) and the phase qubits serving as the target ($Q_{t_l}$)
as inputs. The effect of this operation is modelled as a symbolic phase rotations applied to the target qubits of $U$ ($Q_{t_l}$).

If the control qubit ($q_c$) is in state $\ket{0}$, the target qubit remains unchanged. When $q_c$ is in state $\ket{1}$, the rotational component of each target qubit ($r_{t_l}$) is updated. Formally, this update is represented as
$r^o_{t_l} = r^i_{t_l} +_2 r_l$, 
where $r_l$ denotes the rotation applied by $U$ to the $l^{th}$ phase qubit.

Successive applications of the controlled unitary correspond to increasing powers of $U$: a single application represents $U^{2^0}$, two applications correspond to $U^{2^1}$, and this pattern continues up to $U^{2^{n-1}}$. In this abstraction, only the rotational component is updated, while the basis, superposition, and measurement components of the qubit state remain unchanged.\\

\noindent
\textbf{Definition 5.} \textit{(Abstract Measurement (M-Gate))}  
\textit{
If $m^{i} = \text{2b00}$, then $m^{o} = \text{2b01}$;  
elseif $m^{i} = \text{2b01}$, then $m^{o} = \text{2b10}$;  
elseif $m^{i} = \text{2b10}$, then $m^{o} = \text{2b11}$;  
elseif $m^{i} = \text{2b11}$, then $m^{o} = \text{2b10}$. \\ 
}

\noindent
The measurement abstraction tracks whether a qubit has been measured and enables detection of unintended or repeated measurements. 
A value $m=\text{2b00}$ denotes an unmeasured qubit, while $m=\text{2b01}$ represents a valid single measurement. 
All other values indicate multiple measurement events and are treated as errors (e.g., $m=\text{2b10}$).

Together, these abstractions reduce quantum circuit verification from complex-valued Hilbert space to quantifier-free bit-vector space. 
The corresponding abstraction aligned correctness properties are defined next.

\section{Correctness Properties}
\label{sec:properties}
In this section, we give the abstraction-aligned correctness properties.
The following convention is used for the abstract qubit components in the QPE circuit. The superscript represents the time evolution of qubit states and the subscript represents the qubit number. For example, $q^i_j$, $q^1_j$, $q^{o\text{-}}_j$, $q^o_j$, represents the basis state component at input, first, stage before output and at the output respectively for the $j^{th}$ qubit. Similarly for all other components $s, r, \text{and}\ m$.\\

\noindent
\textbf{Property 1.} \textit{(Superposition Correctness)}
\textit{$\forall \ Q_1, Q_2, \dots , Q_n$ in precision block of QPE circuit, 
$\bigl(s^i_{1}\text{=}$3b000 $\wedge$ $s^i_{2}\text{=}$3b000 $\wedge \dots \wedge$ $s^i_n\text{=}$3b000$\bigr)$
$\rightarrow$
$\bigl(s^1_{1}\text{=}$3b001 $\wedge$ $s^1_{2}\text{=}$3b001 $\wedge \dots \wedge$ $s^1_n\text{=}$3b001$\bigr)$ 
$\bigwedge$ 
$\bigl($$s^o_{1}\text{=}$3b010 $\wedge$ $s^o_{2}\text{=}$3b010 $\wedge \dots \wedge$ $s^o_{n}\text{=}$3b010$\bigr). \quad And, \quad \forall \ Q_1, Q_2, \ldots, Q_p$ in phase block of QPE circuit, $\bigl(s^o = s^i\bigr)$\\
}

\noindent
This property ensures that qubit superposition evolves correctly. 
Precision qubits start outside superposition ($s^i=3b000$). 
The initial $H$-gate creates superposition ($s^1=3b001$), and the inverse QFT (iQFT) returns them to a non-superposition state ($s^o=3b010$), consistent with $H^\dagger=H$. 
Phase qubits are only affected by controlled modular exponentiation (C-U$^{2^k}$), which leaves their superposition unchanged. 
Hence, for all phase qubits, $s^o = s^i$, ensuring superposition correctness across the QPE circuit.\\

\noindent
\textbf{Property 2.} \textit{(iQFT Correctness)}
\textit{$\bigl\langle(\forall  b_1, b_2, \dots, b_n \in \{$\textit1b0, 1b1$\}$:
$\bigwedge_{j=1}^{n}$ $\bigl(r_j^1 \text{=}$ $nbb_jb_{j+1} \ldots b_{n}0\ldots0$  $\wedge \ s^1_j\text{=}$3b001 $\bigr)$
$\rightarrow$
$\bigl(\bigwedge_{j=1}^{n} r^o_j=nb00\dots0 \ \wedge \  q^o_j=b_j \bigr)\bigr\rangle$.\\
}

\noindent
This property verifies the correctness of the iQFT block in QPE~\cite{iqft_arun}. 
The iQFT reverses the rotations applied during the QFT, restoring the qubits to their original states. 
Formally, if the $j^{th}$ input qubit has rotational component $r^i_j = nbb_jb_{j+1}\dots b_n0\dots0$ with superposition $s=3b001$, then after iQFT, the rotation is reset to $r^o_j = nb00\dots0$, and the output basis state matches $q^o_j = b_j$. 
This guarantees that the inverse transformation correctly recovers the encoded bit-vector state for all precision qubits.\\

\noindent
\textbf{Property 3.} \textit{(Measurement Correctness)}
\textit{$\bigl(m^i_{1}\text{=}$2b00 $\wedge$ $m^i_{2}\text{=}$2b00 $\wedge \dots \wedge$ $m^i_n\text{=}$2b00$\bigr) \bigwedge \bigl(m^i_p\text{=}$2b00$\bigr)$
$\rightarrow$
$\bigl(m^{o\text{-}}_{1}\text{=}$2b00 $\wedge$ $m^{o\text{-}}_{2}\text{=}$2b00 $\wedge \dots \wedge$ $m^{o\text{-}}_n\text{=}$2b00$\bigr) $ 
$\bigwedge$
$\bigl(m^o_1$=2b01 $\wedge$ $m^o_2$=2b01 $\wedge \dots \wedge$ $m^o_n$=2b01$\bigr) \bigwedge \bigl(m^i_p$=2b00$)\bigr\rangle$ \\
}

\noindent
This property ensures correct measurement behaviour in QPE. 
Precision qubits must remain unmeasured throughout the circuit until the final output stage: $m^i_1=\dots=m^i_n=2b00$ and $m^{o-}_1=\dots=m^{o-}_n=2b00$. 
At the output, each precision qubit is measured exactly once, $m^o_1=\dots=m^o_n=2b01$. 
Phase qubits are not measured, maintaining $m^i_p=m^o_p=2b00$. 
This guarantees that measurements occur only at the designated output stage, preventing redundant or premature measurement operations.\\

\noindent
\textbf{Property 4.} \textit{(Phase Qubit Correctness)}
\textit{$\forall \ Q_1, Q_2, \dots , Q_n$ in precision block and $Q_{p_1}, Q_{p_1}, \dots , Q_{p_p}$ in the phase block of QPE, the following should hold: \\
$\bigl\langle \forall \ r^i_l\text{= pb00...0},\ l \ \in \{1,2, \dots , p\}$::\\
$(r^o_{p_1} = q_1 \cdot r_1 + q_2 \cdot (2r_1) +  q_3 \cdot (4r_1) + ... + q_n \cdot (2^{n-1} r_1)) \ \wedge$
$(r^o_{p_2} = q_1 \cdot r_2 + q_2 \cdot (2r_2) +  q_3 \cdot (4r_2) + ... + q_n \cdot (2^{n-1} r_2)) \ \wedge$ 
\begin{center}
   $ \vdots $
\end{center} 
$(r^o_{p_p} = q_1 \cdot r_p + q_2 \cdot (2r_p) +  q_3 \cdot (4r_p) + ... + q_n \cdot (2^{n-1} r_p)\bigr)\bigr\rangle$, where $r_1, r_2, \dots , r_p$ are the symbolic rotations applied by the unitary operator U on qubit $Q_{p_1}, Q_{p_1}, \dots , Q_{p_p}$ respectively. \\ 
}

\noindent
Property~4 establishes the functional correctness of the phase accumulation block in the Quantum Phase Estimation (QPE) circuit. 
Let $r_l$ (i.e., $r_1,\dots,r_p$) represent the symbolic rotation angles applied by the unitary operator $U$ as a result of the controlled-$U$ (C-$U$) operations acting on the phase qubits $Q_{p_1},\dots,Q_{p_p}$. A rotation is applied to a phase qubit $Q_{p_k}$ only when its associated control qubit $Q_j$ is in the state $\ket{1}$, in accordance with Definition~4. 

Controlled modular exponentiation is implemented by applying the C-$U$ operation $2^k$ times. Thus, the operation $C$-$U^{2^k}$ can be abstracted as a weighted rotation expressed as $q_k \cdot 2^{k-1} \cdot r_l$, where the resulting rotations are accumulated on each phase qubit using modular arithmetic for all $l \in \{1,2,\dots,p\}\ \text{and} \ k \in \{1,2,\dots,n\}$. 
Satisfying these conditions for every phase qubit ensures that the QPE circuit performs the correct phase rotation.

\section{Verification Methodology}
\label{sec:method}
As discussed in Section~\ref{sec:background}, our verification methodology proceeds in two steps: (i) precision qubit verification, and (ii) phase qubit verification. 
Figure~\ref{fig:ckt_fig_qpe}(c) illustrates a QPE circuit with three precision qubits ($q_0, q_1, q_2$) and one phase qubit ($q_3$), implemented in IBM Qiskit~\cite{qiskit}. 
Modular exponentiation is realized by applying $\text{C-}U^{2^0}$, $2^i$ times on the phase qubit for each precision qubit $i \in \{0,1,2\}$.

The circuit is first abstracted from Hilbert space to the bit-vector domain (Section~\ref{sec:abs}) and encoded into SMT form. 
Verification is performed using the Z3 solver~\cite{z3_cite}, leveraging the four correctness properties. 
Properties~1-3 validate superposition, iQFT, and measurement correctness, while Property~4 ensures correct phase-qubit rotation under $\text{C-}U^{2^k}$. 
We evaluate scalability on QPE circuits with up to 6 precision qubits and $\text{C-}U^{2^k}$ operators spanning up to 1,024 qubits.

Typical QPE errors include: (i) superposition errors from missing or extra $H/H^\dagger$ gates, (ii) iQFT errors from incorrect parameters in $\text{C-}R_k^\dagger$ gates, (iii) measurement errors such as missing, extra, or unintended measurements, and (iv) incorrect controlled modular exponentiation due to wrong control or target qubits. 
Verification of the unitary operator itself is application dependent and not covered here. We assume correct state preparation and unitary implementation. 
The following lemmas demonstrate that any of these errors violate one or more correctness properties, enabling their detection.\\

\noindent
\textbf{Lemma 1.} \textit{In a QPE circuit, applying zero, one or more than two H-gates on a precision qubit, or any H-gate on a phase qubit, violates Property~1.}
\begin{proof}
Consider a qubit $Q_j$. Property~1 enforces that for a precision qubit, $s^i_j=3b000$ at input and $s^o_j=3b010$ at output, corresponding to two H-gate applications: one to create superposition and one in the iQFT. For phase qubits, $s^i_j=s^o_j=3b000$.

\emph{Case 1: No or one H-gate on a precision qubit:} \\
If no H-gate is applied, $s_j$ remains 3b000 (or 3b001 for one) at output (Definition~2), violating the required $s^o_j$=3b010.

\emph{Case 2: More than two H-gate on a precision qubit:} \\
By Definition~2, $s_j$ increments by 3b001 per H-gate. Applying three or more H-gates yields $s^o_j\ne$3b010 (e.g., 3b011 or higher), violating Property~1.

\emph{Case 3: H-gates on a phase qubit:} \\
If any H-gate is applied to a phase qubit, $s_j$ changes from 3b000, violating the requirement $s^o_j$=3b000.

In all cases, Property~1 is violated.
\end{proof}

\noindent
\textbf{Lemma 2.}  
\textit{If the rotation applied by any C-R$^\dagger_k$ gate is incorrect due to misplacement, wrong control, or wrong target qubit, Property~2 is violated.}
\begin{proof}
Let $Q_j$ be a precision qubit. Property~2 requires
\begin{equation}
r^o_j = nb00\ldots0, \text{ and } q^o_j = b_j.
\end{equation}

\emph{Case 1: Missing C-R$^\dagger_k$ gate.} \\
Without the gate, the intended anti-clockwise rotation on $r_j$ is not applied, giving $r^o_j \neq nb00\ldots0$, violating Property~2.

\emph{Case 2: Extra C-R$^\dagger_k$ gates.} \\
Additional gates cause over-rotation, again yielding $r^o_j \neq nb00\ldots0$, violating Property~2.

\emph{Case 3: Incorrect control or target qubit.} \\
Suppose an C-R$^\dagger_k$ gate uses bit $b_q$ in place of the intended bit $b_p$ as the control (or target). The phase rotation applied becomes
\[
2\pi\!\left(\frac{b_q}{2} + \frac{b_{q+1}}{4} + \cdots + \frac{b_m}{2^{m-q}}\right)
\]
instead of the correct
\[
2\pi\!\left(\frac{b_p}{2} + \frac{b_{p+1}}{4} + \cdots + \frac{b_m}{2^{m-p}}\right).
\]
This mismatch in applied phase rotation in all cases where $q\ne p$ yields
\begin{equation}
r^o_j \neq nb00\ldots0,
\end{equation}
This violates the precise rotation condition enforced by Property~2.
Thus, any deviation in count, placement, or qubit assignment of C-R$^\dagger_k$ gates modifies $r$, causing Property~2 to fail.
\end{proof}

\noindent
\textbf{Lemma 3.}  
\textit{If a measurement happens at any stage before the output for precision qubits, or at any stage in phase qubits, it violates Property~3.}
\begin{proof}
Consider the measurement component of a precision qubit $Q_k$ ($m_k$) and a phase qubit $Q_j$ ($m_j$), with $k \in \{1,\dots,n\}$ and $j \in \{1,\dots,p\}$.  Consider the following cases.

\emph{Case 1: Measurement on a precision qubit before output stage.} \\
If a measurement occurs before the output stage, then by Definition~5, at some intermediate stage $x$, $m^x_k \neq 2\text{b}00$ violating Property~3.

\emph{Case 2: Missing measurement on a precision qubit.} \\
If the output measurement is absent, $m^o_k = 2\text{b}01 \neq 2\text{b}01$, violating Property~3.

\emph{Case 3: Accidental measurement of a phase qubit.} \\
If $Q_j$ is measured, $m^o_j \neq 2\text{b}00$, violating Property~3.

Thus, any deviation in measurement operations causes Property~3 to be violated.
\end{proof}

\noindent
\textbf{Lemma 4.}  
\textit{Any error in the controlled-unitary (C-$U$) operation, such as missing C-$U$ gate, additional C-$U$ gate, incorrect control or target assignment of C-$U$ gate, results in an incorrect symbolic rotation on the phase qubit, violating Property~4.}
\begin{proof}
Let $Q_j$ be a control (precision) qubit and $Q_{p_l}$ a phase qubit, Property~4 requires
\begin{equation}
    r^o_{p_l} = \sum_{j=0}^{n-1} 2^{\,j} \cdot q_j \cdot r_l
\end{equation}
Consider the following cases.

\emph{Case 1: Missing or additional C-$U$ gate.} \\
If in a phase qubit $Q_{p_j}$, a C-$U$ gate is missing or there is an additional C-$U$ gate, the corresponding rotational component at the output $r^o_j$ will be different from
what is given in Equation 8, violating Property 4. 
\begin{equation}
\widetilde{r}^o_{p_l} = \sum_{j=0}^{n-1} 2^{j} \cdot q_{j} \cdot r_{l} \pm q_{j} \cdot r_l \neq r^o_{p_l} 
\end{equation}

\emph{Case 2: Incorrect control assignment.} \\
If a wrong control $q_{j'}$ is used, the output becomes
\begin{equation}
\widetilde{r}^o_{p_l} = \sum_{j=0}^{n-1} 2^{j} \cdot q_{j} \cdot r_{l} - q_j\cdot r_l + q_{j'} \cdot r_l \neq r^o_{p_l} \text{when} \ j'\ne j,
\end{equation}
violating Property~4.

\emph{Case 3: Incorrect target assignment.} \\
If the rotation intended for $Q_{p_l}$ is applied to $Q_{p_{l'}}$ ($l'\neq l$), then Property 4 is violated:
\begin{equation}
\widetilde{r}^o_{p_l} = \sum_{j=0}^{n-1} 2^{j} \cdot q_{j} \cdot r_{l} + q_{j} \cdot r_{l'} - q_{j} \cdot r_{l} \neq r^o_{p_l} 
\end{equation}

\emph{Case 4: Combination of cases 1-3.} \\
If an error occurs more than once or a combination of errors from cases 1-3 occurs, the output will be a sum of the error terms in each case. Any such sum will not match the
correct rotation required in Equation 8 and will violate Property 4. \\
\end{proof} 

\noindent
\textbf{Theorem 1.}  
\textit{If the abstracted QPE circuit satisfies Properties~1-4, the corresponding quantum circuit correctly implements the Quantum Phase Estimation algorithm.}
\begin{proof}
A QPE circuit consists of $H$ gates, C-R$^\dagger_k$ gates, C-$U$ gates and $M$ gates. Errors occurring due to missing, misplaced, or additional $H$, C-R$^\dagger_k$, $M$, and C-$U$  gates will be flagged by Properties 1, 2, 3, and 4, respectively, as shown in Lemmas~1, 2, 3, and 4, respectively. 
$H$ gate errors are flagged by errors in the superposition component. 
$M$ gate errors are flagged by the measurement component. 
C-R$^\dagger_k$ gate errors are flagged by the rotational component
and the basis component of the precision qubits. 
C-$U$ gate errors are flagged by the rotational component of the phase qubits. 
Thus, the four error classes impact different abstract components.
Therefore, any combination of errors will be flagged accurately and 
detection of errors in one class will not interfere with errors in 
another class. 
\end{proof}

\section{Experimental Results}
\label{sec:results}
This section presents the verification benchmarking results.  
All experiments were performed on a system with an Intel\textregistered~Core\texttrademark~Ultra~9~285K CPU at 3.2~GHz, 192~GB RAM, running RHEL~9 (64-bit), using Z3 SMT solver v4.8.12.  
Verification proceeds in two stages as described in Section~\ref{sec:method}.  
The iQFT stage is verified using the method in~\cite{iqft_arun}. For a 6-qubit precision configuration, verification completes in 0.01~s with peak memory of 17~MB. A 6-bit precision provides phase accuracy up to $1/2^6$.  
\begin{table}[!ht]
\caption{Phase Block Verification Results}
\begin{center}
\resizebox{.48\textwidth}{!}{
\begin{tabular}{c c c c c}
\toprule
    \textbf{Circuit Benchmark} &
    \multicolumn{2}{c} {\textbf{Correct Circuit}} &
    \multicolumn{2}{c} {\textbf{Error Circuit}} \\ 
    \textbf{Phase Qubits Count}&
    \textbf{Time}(s) &
    \textbf{Memory}(MB) &
    \textbf{Time}(s) &
    \textbf{Memory}(MB) \\    
\midrule
    2 & 0.33 & 27.4 & 0.03 & 27.4 \\
    4 & 0.91 & 29.1 & 0.09 & 29.1 \\ 
    8  & 1.91 & 40.7 & 0.09 & 40.74 \\
    16  & 7.42 & 80.3 & 0.41 & 80.3 \\
    32  & 45.73 & 149 & 1.10 & 150 \\
    64  & 196 & 243 & 0.80 & 292 \\
    128 & 1,407 & 480 & 15.4 & 481 \\
    256 & 10,524 & 1,108 & 142 & 872 \\ 
    512 & 64,383 & 3,109 & 8.05 & 1,717 \\
    1,024 & 288,196 & 7,663 & 15.9 & 3,434 \\ 
\bottomrule 
\label{tab:phase_block_results}
\end{tabular}
}
\end{center}
\end{table}
Table~\ref{tab:phase_block_results} summarizes results for 6-bit precision circuits with phase-qubit counts from 2 to 1,024.  
The \emph{Phase Qubits Count} column indicates the number of phase qubits.  
\emph{Time (s)} and \emph{Memory (MB)} report total verification time and peak memory, respectively.  
\emph{Correct Circuit} corresponds to error-free C-$U^{2^k}$ implementations, while the \emph{Error} columns show results for injected control-qubit errors at the first qubit. Other error scenarios are verified and consume fewer resources than correct circuit and are omitted from reporting for brevity..  
As observed, verification time for larger circuits with errors (512, 1024 qubits) is lower despite higher memory usage, due to internal Boolean optimizations in z3.
\section{Conclusion and Future Works}
\label{sec:conclusion}
We demonstrated a symbolic abstraction framework using quantifier-free bit-vector logic to model qubits, quantum gates, and operators, extending the abstraction to include measurement along with rotation and superposition.  
This framework reduces the verification of Quantum Phase Estimation (QPE) circuits from the complex-valued Hilbert space to a bit-vector domain.  
We define formal correctness properties capturing the functional behaviour of QPE circuits and lemmas covering key error scenarios, ensuring that all such errors are detectable.  
Experimental results demonstrate scalability, successfully verifying QPE circuits with up to 1,024 phase qubits in the unitary operator and 6-bit precision qubits.  
Future work will focus on reducing verification time for larger circuits and extending the framework to formally verify concrete unitary operators, enabling analysis of more complex quantum algorithms.

\section*{Acknowledgments}

This paper is based upon work supported by the National
Science Foundation under Grant No. 2513276.

\bibliographystyle{IEEEtran}
\bibliography{ref.bib}

\end{document}